\documentclass[letterpaper, 10 pt, conference]{ieeeconf}  % Comment this line out if you need a4paper

\IEEEoverridecommandlockouts                              % This command is only needed if 
\usepackage{graphics} % for pdf, bitmapped graphics files
\usepackage{epsfig} % for postscript graphics files
\usepackage{mathptmx} % assumes new font selection scheme installed
\usepackage{times} % assumes new font selection scheme installed
\usepackage{amsmath} % assumes amsmath package installed
\usepackage{amssymb}  % assumes amsmath package installed
\usepackage{pifont}

\title{\LARGE \bf
Magnetic-Guided Flexible Origami Robot  toward Long-Term Phototherapy of H. pylori in the Stomach}

\author{Sishen Yuan$^{1,3}$, Baijia Liang$^{1}$, Po Wa Wong$^{1}$, Mingjing Xu$^{1}$, Chi Hsuan Li$^{1}$, Zhen Li$^{2}$, Hongliang Ren$^{1,3*}$% <-this % stops a space 
\thanks{The work was supported by NSFC/RGC Joint Research Scheme N\_CUHK420/22; Hong Kong Research Grants Council (RGC) Collaborative Research Fund (CRF C4026-21GF and CRF C4063-18G), and General Research Fund (GRF 14203323, GRF 14216022, and GRF 14211420); Shenzhen-Hong Kong-Macau Technology Research Programme (Type C) Grant 202108233000303; CUHK-SZRI grant: the key project 2021B1515120035 (B.02.21.00101) of the Regional Joint Fund Project of the Basic and Applied Research Fund of Guangdong Province. 
(Corresponding author: Hongliang Ren.)}% <-this % stops a space
\thanks{$^{1}$Sishen Yuan, Baijia Liang, Po Wa Wong, Mingjing Xu, Chi Hsuan Li, and Hongliang Ren are with the Department of Electronic Engineering, The Chinese University of Hong Kong, Hong Kong.}%
\thanks{$^{2}$Zhen Li is with the Department of Gastroenterology, Laboratory of Translational Gastroenterology, and Robot Engineering Laboratory for Precise Diagnosis and Therapyof GI Tumor, Qilu Hospital of Shandong University, Cheeloo College of Medicine, Jinan, Shandong, China}
\thanks{$^{3}$Sishen Yuan and Hongliang Ren are with Shenzhen Research Institute, The Chinese University of Hong Kong, Shenzhen 518063 China}
}

\begin{document}

\maketitle
\thispagestyle{empty}
\pagestyle{empty}

%%%%%%%%%%%%%%%%%%%%%%%%%%%%%%%%%%%%%%%%%%%%%%%%%%%%%%%%%%%%%%%%%%%%%%%%%%%%%%%%
\begin{abstract}

Helicobacter pylori, a pervasive bacterial infection associated with gastrointestinal disorders such as gastritis, peptic ulcer disease, and gastric cancer, impacts approximately 50\% of the global population. The efficacy of standard clinical eradication therapies is diminishing due to the rise of antibiotic-resistant strains, necessitating alternative treatment strategies. Photodynamic therapy (PDT) emerges as a promising prospect in this context.
This study presents the development and implementation of a magnetically-guided origami robot, incorporating flexible printed circuit units for sustained and stable phototherapy of Helicobacter pylori. Each integrated unit is equipped with wireless charging capabilities, producing an optimal power output that can concurrently illuminate up to 15 LEDs at their maximum intensity. Crucially, these units can be remotely manipulated via a magnetic field, facilitating both translational and rotational movements.
We propose an open-loop manual control sequence that allows the formation of a stable, compliant triangular structure through the interaction of internal magnets. This adaptable configuration is uniquely designed to withstand the dynamic squeezing environment prevalent in real-world gastric applications. The research herein represents a significant stride in leveraging technology for innovative medical solutions, particularly in the management of antibiotic-resistant Helicobacter pylori infections.

\end{abstract}

%%%%%%%%%%%%%%%%%%%%%%%%%%%%%%%%%%%%%%%%%%%%%%%%%%%%%%%%%%%%%%%%%%%%%%%%%%%%%%%%
\section{Introduction}

Helicobacter pylori, a bacterial pathogen ubiquitously present, pervades roughly half of the global populace, its incidence notably escalated in economically emergent nations \cite{azevedo2009epidemiology,hooi2017global}. 
This pervasive pathogen is implicated as a primary aetiological agent in a variety of gastrointestinal pathologies encompassing gastritis, peptic ulcer disease, and gastric cancer \cite{cover2009helicobacter,graham2014history}. Alarmingly, H. pylori infection is attributable for an estimated 90\% of duodenal ulcers and 80\% of gastric ulcers \cite{sousa2022helicobacter}.
The conventional clinical protocol for H. pylori infection management is a regimen coined as 'triple therapy' \cite{liou2020treatment,chey2017acg}. This typically encompasses a proton pump inhibitor (PPI) and a pair of antibiotics - most frequently clarithromycin and either amoxicillin or metronidazole. Generally, this regimen is dispensed over a span of 10 - 14 days. Despite its vast implementation, the efficacy of this therapeutic approach is becoming increasingly undermined due to the mounting incidence of antibiotic resistance, most notably towards clarithromycin.
%As an alternative strategy, the administration of quadruple therapy, which includes a PPI, bismuth, tetracycline, and a nitroimidazole antibiotic, is often pursued in instances where triple therapy has been unsuccessful, or when it is anticipated to be less effective due to established antibiotic resistance profiles. Concurrently, innovative therapeutic strategies are subjects of rigorous investigation, such as probiotics, vaccines, and bacteriophages, with the overarching goal of augmenting eradication rates and attenuating side effects associated with existing treatment regimens.

Photodynamic therapy (PDT) represents a potential therapeutic avenue in the combat against Helicobacter pylori infection. This methodology entails the use of a photosensitizer, activated by light of a specified wavelength, culminating in the production of reactive oxygen species capable of inducing damage to bacterial cells \cite{hamblin2005helicobacter,battisti2017spectroscopic}. Intriguingly, H. pylori demonstrates an innate ability to synthesize and accumulate the photosensitizer, porphyrin, rendering it an apt candidate for antimicrobial PDT, thereby eliminating the need for an external photosensitizer \cite{hamblin2005helicobacter}. In vitro and in vivo animal models have exhibited effective bactericidal action of PDT against H. pylori \cite{li2016capsule}. PDT presents several advantages over conventional antibiotic therapy, including the capacity for cell-specific targeting and the potential to overcome the hurdle of antibiotic resistance. Moreover, PDT has been associated with a reduced risk of side effects in comparison to antibiotic therapy. Notwithstanding, PDT's clinical application for H. pylori infection is in its nascent stages, necessitating further research to ascertain its safety and efficacy in human trials. For instance, PDT's therapeutic effect is time-dependent, with H. pylori repopulating after a brief PDT period \cite{li2016capsule}. 

%PDT's challenges in treating H. pylori infection include the requirement for efficacious photosensitizers that can specifically target the bacterium and the development of optimal light delivery systems that can successfully reach the infection site within the stomach. Consequently, this manuscript will concentrate on addressing the latter challenge, exploring methods to establish a long-term stable light delivery system.
%%
\begin{figure*}
    \centering
    \includegraphics[width=1.0\linewidth]{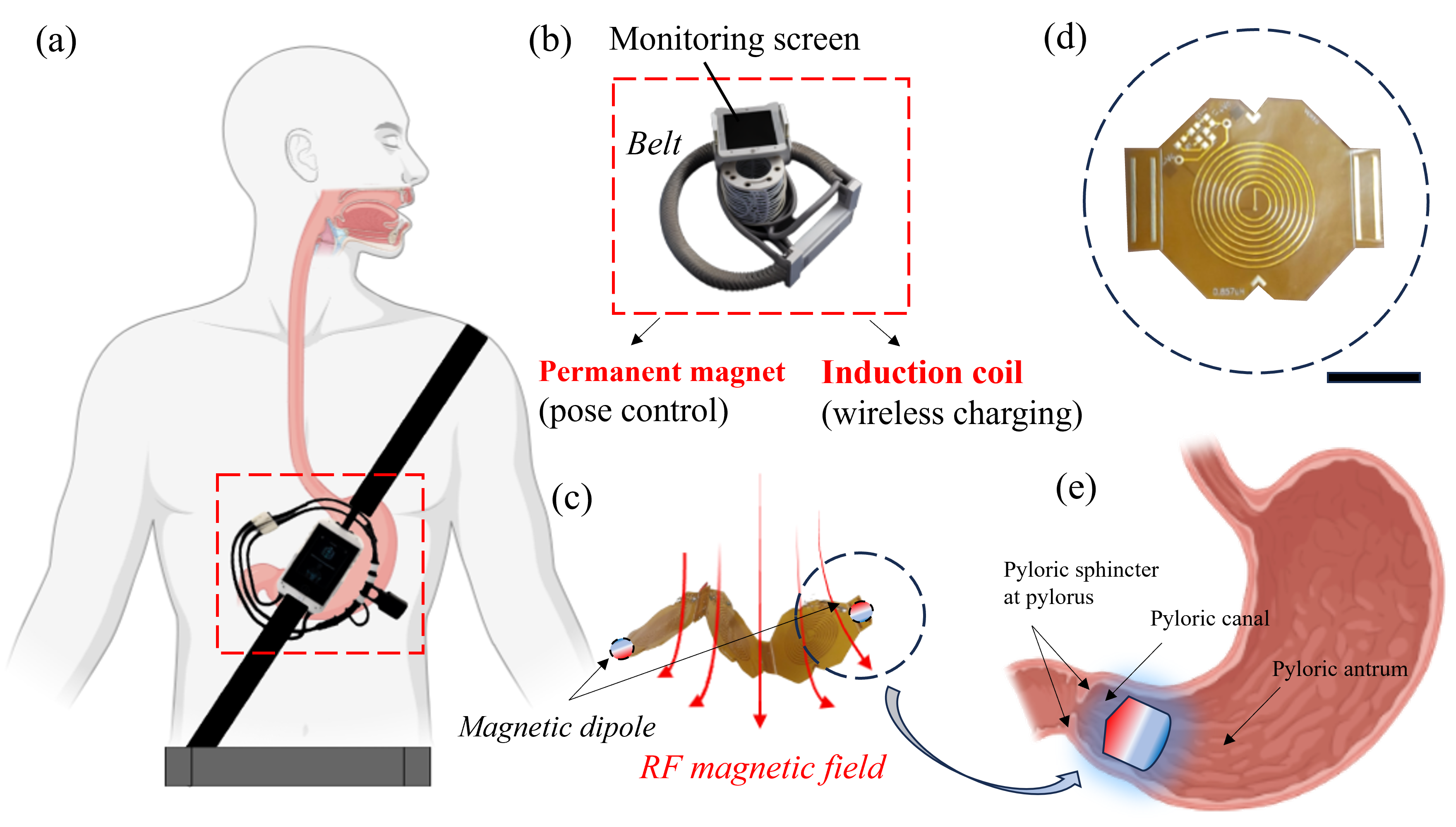}
    \caption{Long-term Deployment System for Helicobacter pylori Photodynamic Treatment. (a) System overview showing an external device for possible intragastric environment detection, MGOR pose control, and wireless energy provision. (b) The external wearable device visualization schematic shall contain the external permanent magnet and the induction coil. (c) Illustration of the pose control mechanism of the MGOR, facilitated by permanent magnets and a spiral copper coil for wireless charging. (d) Flex-printed circuit board (F-PCB) unit. Scale bar: 1 cm. (e) Deployment strategy of the MGOR within the pyloric canal to minimize movement towards the pyloric antrum, a region conducive for H. pylori survival and replication.}
\end{figure*}

Origami engineering enhances biomedical devices with compact, expandable designs for less invasive procedures and improves human-machine interfaces with ergonomic adaptability \cite{cai2023magnetically,yeow2022magnetically,ze2022soft,salerno2016novel,zhang2023active,wu2021multifunctional}, notably within the digestive tract. This is largely attributed to their capacity to enact a variety of motions under actuation, coupled with their ability to adaptively deform \cite{yuan2022versatile}. These pliable structures exhibit compliance with their ambient environment, marking them as particularly suitable for the highly variable and dynamic conditions of the gastrointestinal tract \cite{miyashita2016ingestible}.
Among the diverse technologies contributing to the multifunctional integration of origami structures, Printed Circuit Board (PCB) technology plays a pivotal role. It furnishes the requisite conduit for interaction and transmission of signals encompassing various modalities \cite{wu2021multifunctional,chen2022origami,lee2018origami,li2021miura}. 
This insight prompts the envisagement of a class of origami-based robotic constructs, built upon flex-printed circuit board (F-PCB) units, enabling the wireless transfer and conversion of energy from electrical to luminous form \cite{ho2014wireless}.

In this paper, we present the implementation of flexible printed circuit units, integrated into a magnetically-guided origami robot (MGOR), designed for stable light delivery in the long-term phototherapy of Helicobacter pylori. Each F-PCB unit possesses the capability of wireless charging, with the ideal power output capable of illuminating 15 LEDs simultaneously to their maximum intensity. Importantly, these units can be wirelessly manipulated via a magnetic field for both translational and rotational movement. Furthermore, the proposed open-loop manual control sequence enables the formation of a stable triangular compliant structure, facilitated by the interaction of internal magnets. This adaptive structure is designed to accommodate the dynamic squeezing environment prevalent in real-world applications, as shown in Fig. 1.

%% Part 2

\section{Delivery, Magnetic Actuation and Deployment}

\begin{figure}
    \centering
    \includegraphics[width=1.0\linewidth]{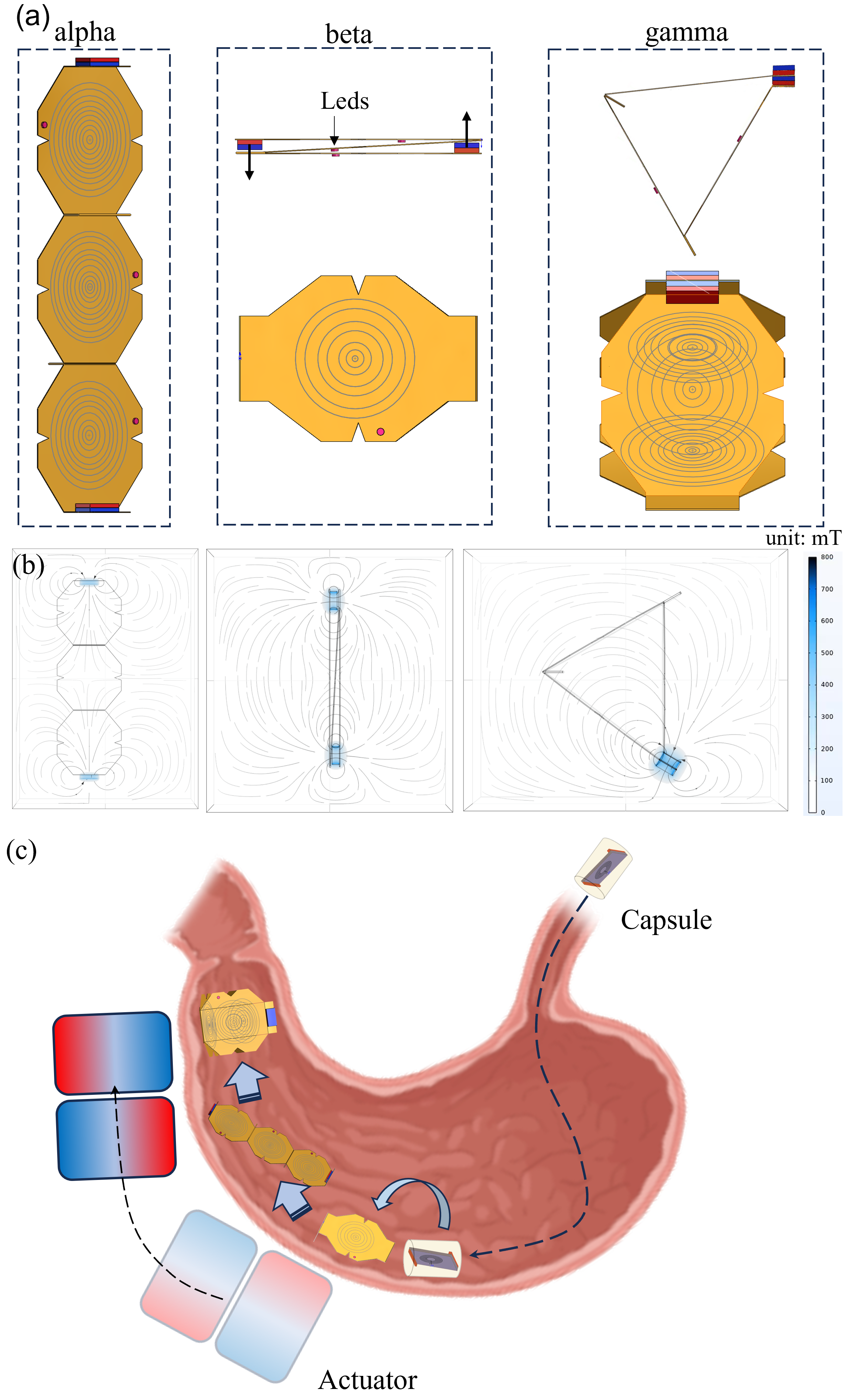}
    \caption{MGOR design and state transitions. (a) Design of the MGOR composed of three F-PCB cells with IPMs, stabilized in three distinct states (alpha, beta, and gamma) using an external magnetic field. (b) COMSOL simulations of the magnetic field distribution of the IPMs, highlighting fluctuations in internal forces during state transitions. (c) Demonstration of MGOR in its folded state, encapsulated within a capsule casing for oral administration, and its subsequent unfolding into the beta state.}
\end{figure}

A conspicuous challenge associated with the morphology self-locking and wireless actuation of flexible structures, predicated on internal magnetic interactions, lies in the limited accessibility of the morphology self-locking as a localized energy steady state. This limitation can lead to the unintended generation of a steady state due to potential disturbances in the wireless actuation process.
Building upon our prior research \cite{yuan2022versatile,swaminathan2021multiphysics}, we present a design for the MGOR comprising three F-PCB cells, fitted with two small rectangular internal permanent magnets (IPMs: 10 mm length, 5 mm width, 2 mm height) on either end. The orientation of the magnetic poles is depicted in Fig. 2(a). The chosen material for the body is polyethylene terephthalate with a thickness of 0.1 mm. Each F-PCB cell is notched to imbue it with the capacity to accommodate deformation.
Consequently, the MGOR, as conjectured, can be stabilized in three distinct states (labeled as alpha, beta, and gamma) with the application of an external magnetic field, as depicted in Fig. 2(a).
The magnetic field distribution of the IPMs has been visualized within the COMSOL simulation environment, as depicted in Fig. 2(b). The observed fluctuations – both decay and enhancement – in the internal forces during typical state transitions form a critical element in imparting the MGOR with a diverse and stable morphology.

We leverage the elastic stiffness of the MGOR material to offset the internal magnetic attraction in its folded state. It is proposed that the MGOR can be folded and encapsulated within a capsule casing, which enables patient-friendly oral administration to the stomach, as demonstrated in Fig. 2(c). Thereafter, it naturally unfolds into a beta state. Through manipulation of the magnetic field, the MGOR can be transformed into an alpha state for short-term, targeted phototherapy at specific stomach wall regions, or into a gamma state for deployment within the pyloric canal. Nonetheless, designing an external magnetic field manipulation sequence to actualize the aforementioned morphological transformations presents a significant challenge.

\begin{figure*}
    \centering
    \includegraphics[width=1.0\linewidth]{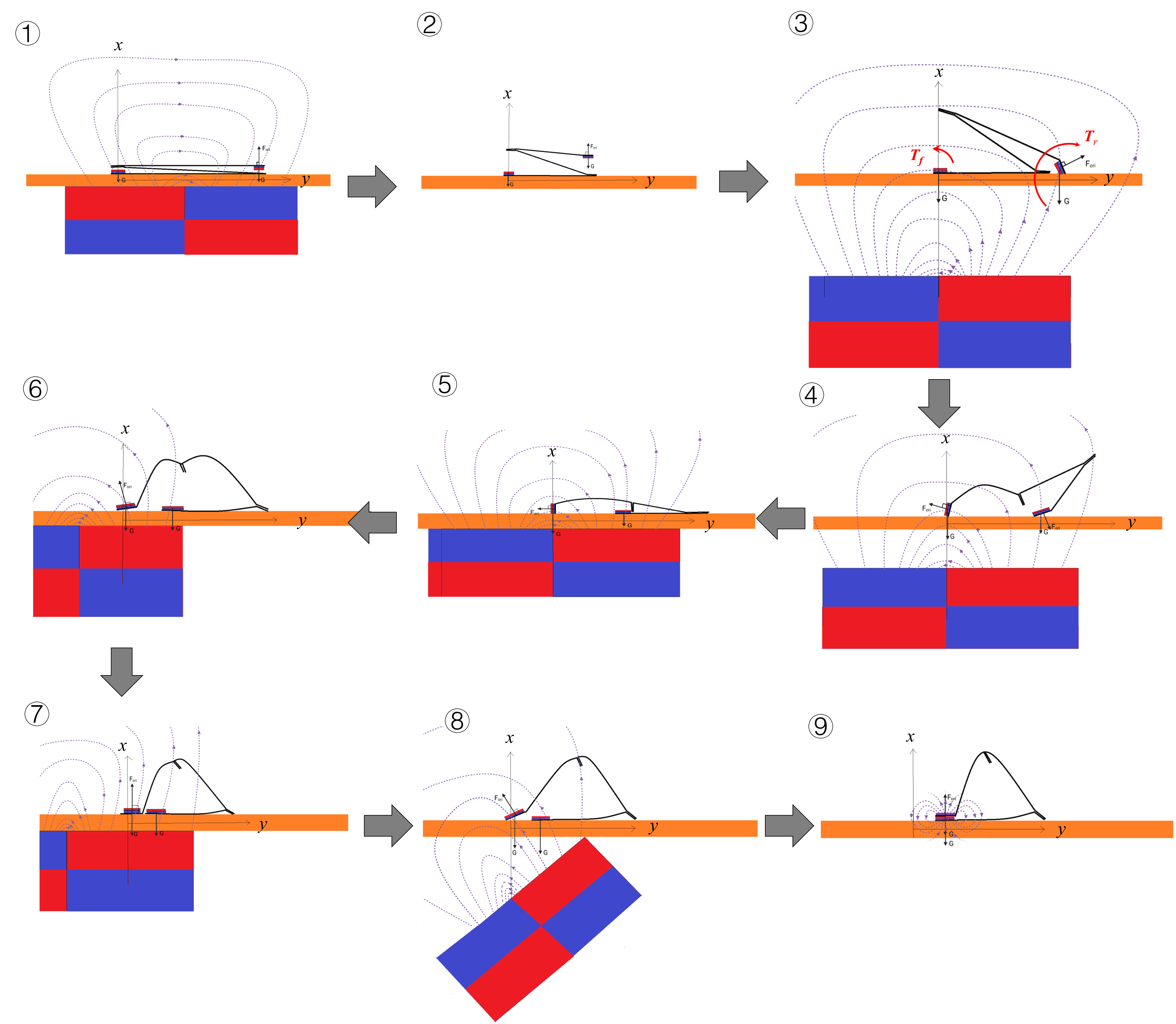}
    \caption{Comprehensive control methodology for MGOR state transition. The figure illustrates the use of an external controller, composed of two axially magnetized cylindrical magnets arranged in parallel, to facilitate the transition of the MGOR from beta to gamma state. The process includes maintaining the MGOR in the beta state, inducing divergent moments on the L-IPM and R-IPM by swift rotation of the EPM around the $x$-axis and linear movement along the $y$-axis, and inducing the flipping of the L-IPM when the IPMs enter an in-plane repulsive state.}
\end{figure*}

Accordingly, Figure 3 elucidates a comprehensive control methodology utilizing an external controller, composed of two axially magnetized cylindrical magnets (diameter: 40 mm, height: 50 mm), arranged in parallel as a single unit. This facilitates the transition of the MGOR from the beta state to the gamma state. Drawing from the aforementioned equations, it is inferred that the left and right IPMs, designated as L-IPM and R-IPM respectively, can self-assemble when guided to approximately $\sim$5 mm apart. \ding{172} demonstrates that the current configuration can maintain the MGOR in the beta state. However, by swiftly rotating the EPM around the $x$-axis, divergent moments can be induced on the L-IPM and the R-IPM, effectively causing the MGOR to flip. Simultaneously, linear movement of the EPM along the $y$-axis creates a magnetic gradient that draws the IPMs closer together. Occasionally, as depicted in \ding{178}, the IPMs may enter an in-plane repulsive state. Consequently, it becomes necessary to rotate the EPM around an axis perpendicular to the paper surface to induce flipping of the L-IPM, ultimately facilitating the autonomous assembly of the IPMs.
%%%%

%The H.p. OR's morphological control is a repeatable sequence of actions resulting from magnetic field manipulation, operating synergistically with the FPC's compliant structure. This sequence is informed by prior work and experiences\cite{swaminathan2021multiphysics}. Establishing a quantitative description of the morphology and mapping the actuator's pose parameters will pave the way for closed-loop visual servo-navigation deployment of the H.p. OR. This remains a future objective, and for now, this study reports on one potential method.
%[HR: future work?]
%

We conducted a physical verification following the motion sequence outlined in Figure 3, the results of which are presented in Figure 4. The keyframes extracted from the operational footage exhibit considerable consistency with the intended morphological plan.

\begin{figure}
    \centering
    \includegraphics[width=1.0\linewidth]{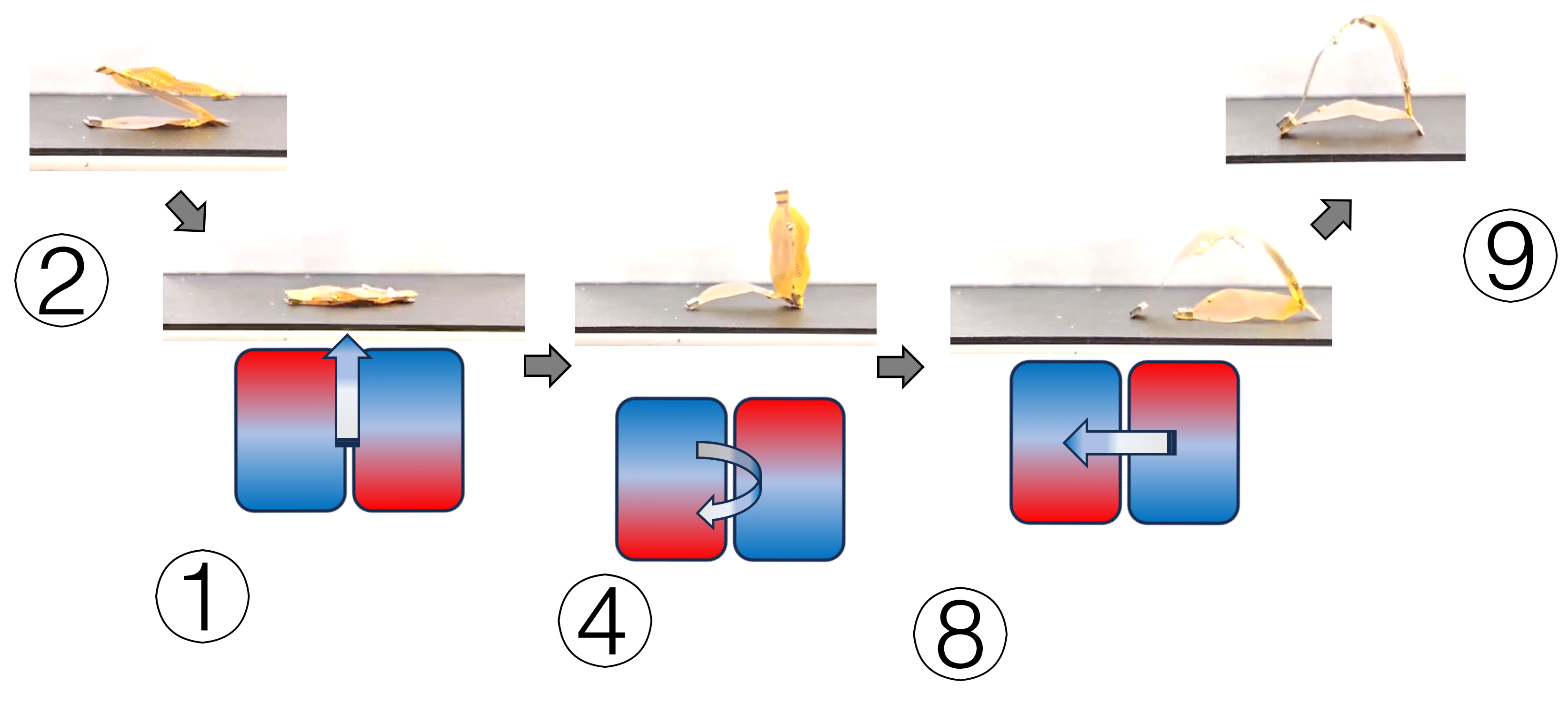}
    \caption{Physical verification of MGOR motion sequence. Keyframes from the operational footage demonstrating notable consistency with the motion sequence outlined in Figure 3.}
\end{figure}

To thoroughly evaluate the robustness of the unfolded state, we manually exerted lateral and longitudinal squeezing forces, ranging from 2 to 10 N, in both vertical directions. The outcomes of the COMSOL simulations and manual experiments are depicted in Figure 5(a) and Figure 5(b) respectively. Upon removal of these applied pressures, the MGOR demonstrated a complete recovery to its initial unfolded state.

\begin{figure}
    \centering
    \includegraphics[width=1.0\linewidth]{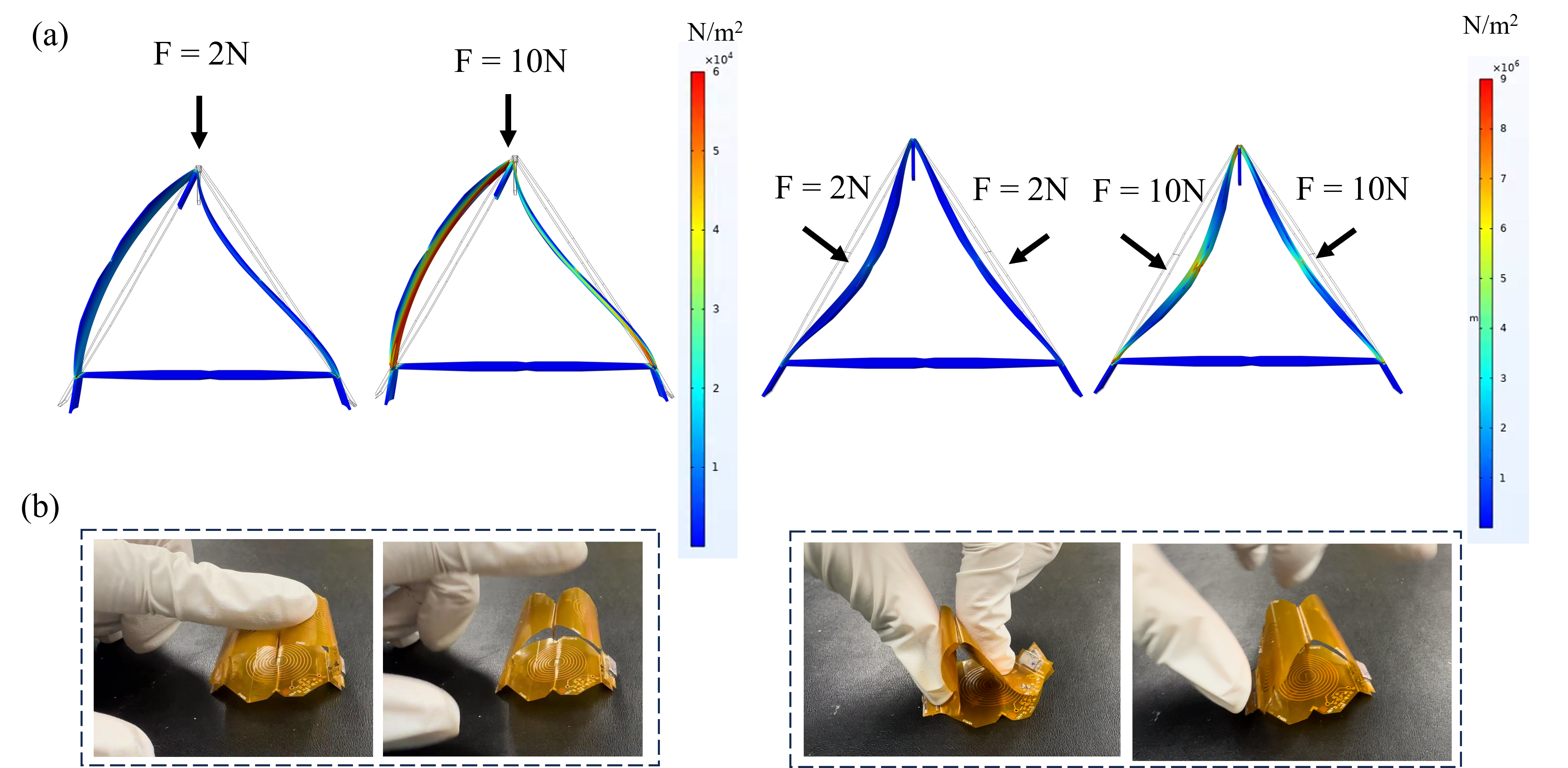}
    \caption{MGOR's robustness under force. (a) COMSOL simulations under 2-10N forces in vertical directions. (b) MGOR's full recovery post-manual force application.}
\end{figure}

\begin{figure*}
    \centering
    \includegraphics[width=1.0\linewidth]{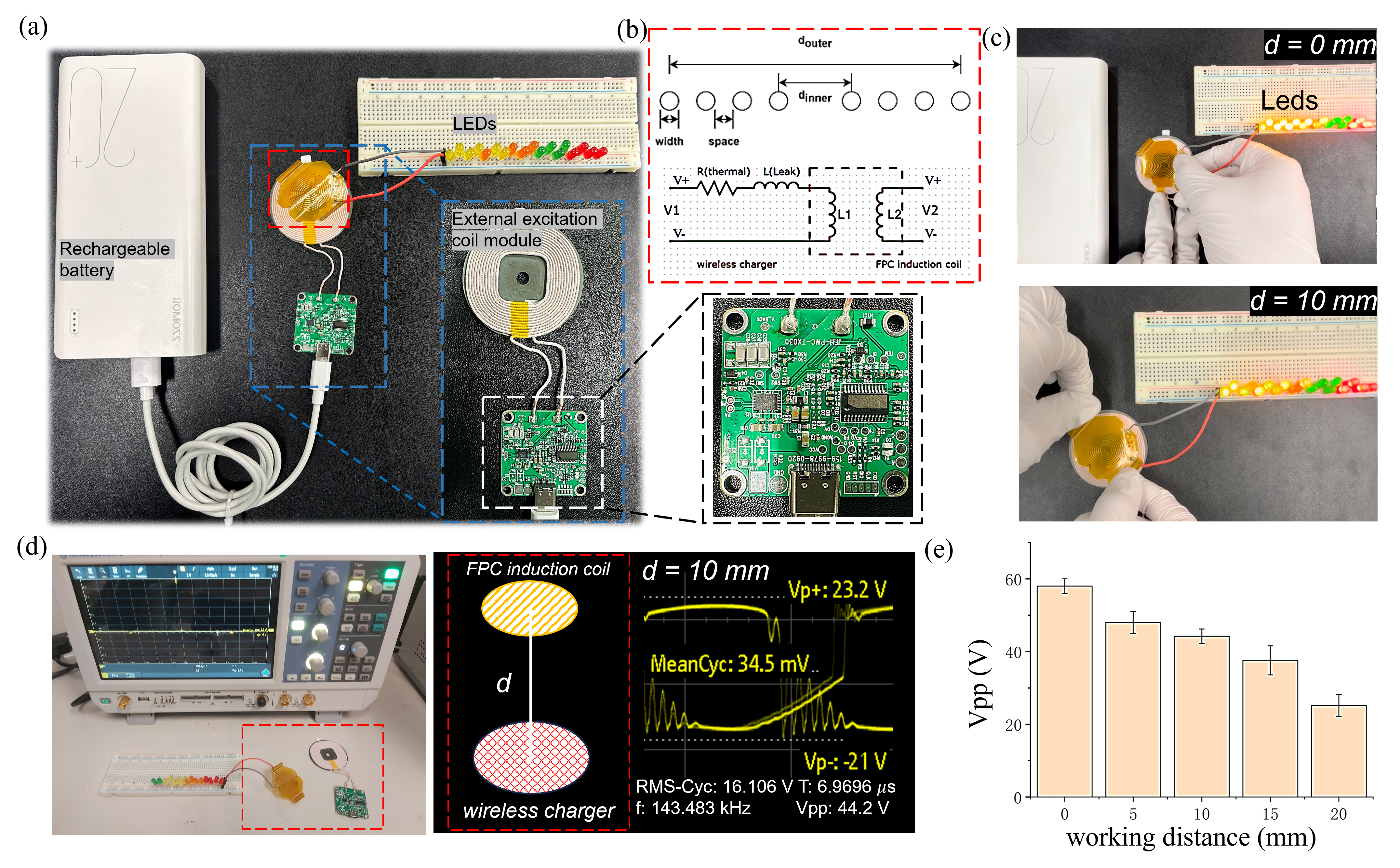}
    \caption{The MGOR wireless charging system and its performance. (a) Depiction of the MGOR wireless charging system, comprising a rechargeable battery, an external excitation coil module, the MGOR, and LEDs for testing. (b) Theoretical framework underlying the double spiral structure (DSS) for voltage amplification, based on Faraday's law of electromagnetic induction. (c) Demonstration of wireless charging capability with an array of 15 LEDs connected in parallel, each consuming 80 mW of power. Brightness is maintained up to a 12 mm separation. (d) Induced voltage and frequency measurements using an oscilloscope. (e) Voltage amplitude decreases from 58 V to 25.2 V as the separation between the wireless charger and the F-PCB induction coil increases from 0 mm to 20 mm. The critical threshold distance of 10 mm corresponds to a voltage of 44.2 V.}
\end{figure*}

\section{F-PCB Circuit Design and Wireless Charging}

Figure 6(a) depicts the wearable MGOR wireless charging system, which consists of several integral components: a rechargeable battery (output: 5 V; 18 W), an external excitation coil module, the MGOR, and LEDs utilised for testing.
The F-PCB is fabricated from two 0.05 mm-thick layers of polyethylene terephthalate (PET), interspersed with a copper trace. To enable power reception via magnetic induction, a spiral copper coil is embedded into a double-sided FPC, a configuration termed the double spiral structure (DSS).

The DSS demonstrates the ability to amplify the voltage gain twofold, an effect explicable by Faraday's law of electromagnetic induction (Equation 3). This principle suggests a direct correlation between the rate of flux change in the magnetic field and the induction coil's area, which in turn influences the induced electromotive force, or voltage. The increased turns of the spiral coil effectively double the induced voltage, highlighting the power amplification capabilities of the DSS.

\begin{equation} \label{eu_eqn}
\varepsilon= -N\frac{\delta\Phi}{\delta t}
\end{equation}

Here, $\varepsilon$ denotes the induced voltage; $N$ signifies the number of turns of a spiral coil; $\delta\Phi$ represents the change in magnetic flux; $\delta t$ indicates the change in time.

To maximize transmission efficiency, our design incorporates surface-mounted capacitors, enabling precise impedance adjustment within the receiver (Rx) circuit. This Rx circuit is condensed to an LC series circuit. The capacitive reactance counteracts the inductive reactance on Rx when the resonant frequencies of the F-PCB align with the transmitter (Tx) emitted electromagnetic wave frequency, and the resonant capacitor minimizes inductive leakage effects. Consequently, power dissipation is reduced, enhancing power efficiency.

The Rx coil's inductance is instrumental in determining the surface-mount capacitor's capacitance for resonance. Using Harold A. Wheeler's Approximation, the single-sided spiral copper coil's inductance is estimated at approximately 0.857 µH, while the double-sided coil's inductance is approximately 1.714 µH.

\begin{equation} \label{eu_eqn}
L(\mu H) = \frac{r^{2} n^{2}} {8r + 11w}
\end{equation}

Here, $r$ signifies the radius to the center of windings; $w$ denotes the width of windings; $n$ represents the number of turns.

Incorporating different capacitance values into Equation (5) allows for adjustable tuning of the resonance frequency on the Rx coil. With this tunable capacitance feature, F-PCB can accommodate different frequency standards of wireless chargers.

\begin{equation} \label{eu_eqn}
f=\frac{1}{2\pi\sqrt{LC}}
\end{equation}

The Tx frequency is approximately 145 kHz. Utilizing the resonant frequency of the LC in-series circuit equation, it is determined that a 1 µF capacitor is suitable for the F-PCB by substituting the Rx inductance and Tx frequency.

Figure 6(c) illustrates the wireless charging capacity of the developed prototype, with a setup of 15 LEDs connected in parallel, each dissipating 80 mW of power. At a proximity of approximately 0 mm between the MGOR and the external excitation coil, all LEDs were fully lit. As the distance between the two components was incrementally increased, illumination was maintained up to a gap of 12 mm, although brightness was reduced.

Subsequent induced voltage and frequency measurements were performed using an oscilloscope (Fig. 6(d)). As the separation $d$ between the wireless charger and the F-PCB induction coil increased progressively from 0 mm to 20 mm, a corresponding decrease in the induction voltage amplitude from 58 V to 25.2 V was observed (Fig. 6(e)).
It is worth noting that prior illumination experiments identified 10 mm as a critical threshold distance, and in this context, the corresponding voltage of 44.2 V serves as a benchmark.

\section{Discussion and Conclusion}

\begin{figure}
    \centering
    \includegraphics[width=1.0\linewidth]{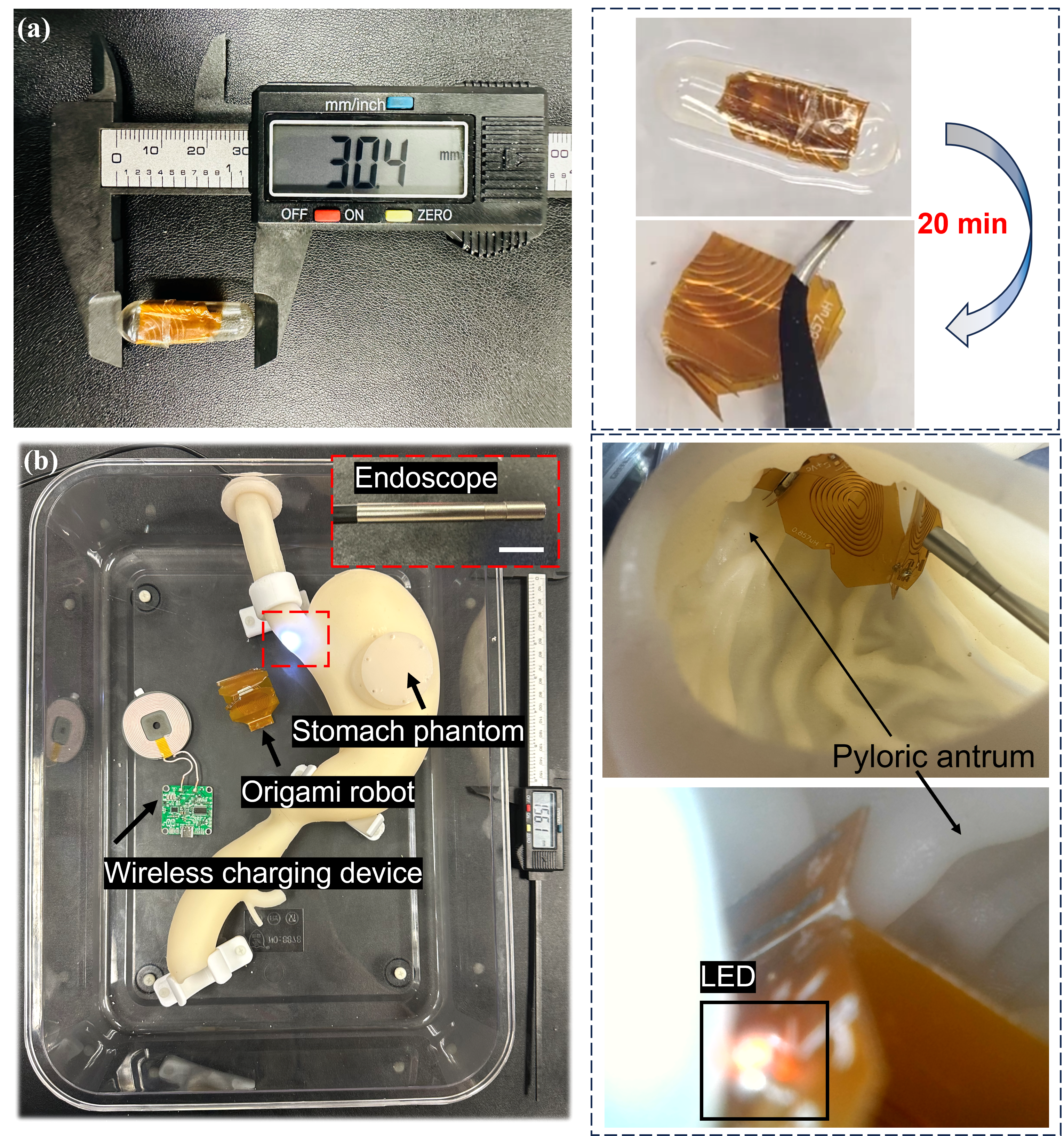}
    \caption{MGOR delivery and operation in a simulated gastric environment. (a) Water-soluble capsule with MGOR, demonstrating dissolution within 20 minutes. (b) MGOR in a gastric phantom, showing size compatibility and functional integration demonstrated by LED flickering upon wireless charging.}
\end{figure}

Figure 7(a) demonstrates the potential of delivering the MGOR into the stomach via a water-soluble capsule. Constructed from glutinous rice, the edible shell of the capsule gradually dissolves when exposed to water. With dimensions of 8 mm in diameter and 30.4 mm in length, the capsule completely dissolves within an estimated 20-minute timeframe.
We also conducted an operational demonstration using a human gastric phantom, as depicted in Fig. 7(b). The MGOR was positioned within the gastric sinus, and an endoscope was inserted into the stomach via the esophagus to observe the MGOR.
As indicated in Fig. 7(b), the dimensions and morphology of the MGOR align well with the physiological structure of the gastric sinus. The positioning of the wireless charging device on the mimic body's surface resulted in noticeable LED flickering, further substantiating the operational integration of the MGOR within the simulated gastric environment.

While this paper delineates the control sequences governing the state transitions of the MGOR with clarity, the evolution towards a fully autonomous control system remains nascent. There is a requisite for advanced development in both MGOR kinematic modeling and morphological perception algorithms. Our previous work leverages established magnetic localization techniques to facilitate non-visual morphological perception via the analysis of magnetic field distributions from integrated position markers \cite{song2021magnetic,su2023amagposenet,su2023magnetic}. However, the current validation of the MGOR’s application is confined to in vitro phantom models, and thus does not encompass in vivo experimentation to assess potential risks such as mucosal damage within the gastric environment. Moreover, the efficiency and capability of the wireless energy transfer system introduced herein warrant further refinement to meet clinical efficacy and safety standards \cite{zhang2023wirelessly}.

%This research presents a novel F-PCB based origami robot, the MGOR, engineered for prolonged deployment within the gastric sinus for phototherapy of Helicobacter pylori. Our results illustrate that wireless charging can effectively power the MGOR for sustained deployment. Additionally, control via a magnetic field shows significant potential for wireless in vivo deployment following capsule ingestion.

%However, certain elements were not explored within this study. These include aspects of biocompatibility, experimental testing of MGOR deployment durations, as well as potential physical harm or chemical toxicity that could stem from the angles and electronic circuitry housed within the origami structure. Moreover, the in vivo gastric environment represents a dynamic and complex context, significantly differing from in vitro conditions. The effectiveness of magnetic control in maneuvering origami structures within such an environment awaits verification.
%These unresolved aspects and the challenges they pose will serve as the foundation for future investigations, as we strive to refine and broaden the potential applications of the MGOR system.

\bibliographystyle{ieeetr}
\bibliography{reference}

\end{document}